\documentstyle[twoside,fleqn,espcrc2,amsfonts,epsfig,graphics]{article}

\def\COMENTARIO#1{}        

\def\bbox#1{{ \mbox{\boldmath $#1$}} }

\def\num#1{\hbox{$^{\rm #1}$}}
\newcommand{\cruza}{\!\not\!}
\def\slash#1{\raise.18ex\hbox{/}\kern-.50em #1}

\title{A new mechanism of mass protection for 
        fermions\thanks{
        Presented by J.~M.~Carmona ({\tt carmona@difi.unipi.it}).
        Partially supported by CICyT (Spain) \hbox{AEN97-1680}, 
        AEN97-1708, and TMR Network no. \hbox{FMRX-CT97-0122}.
}}

\author{J.L.~Alonso\address{Departamento de F\'{\i}sica Te\'orica,
        Facultad de Ciencias, Univ. de Zaragoza, 50009 Zaragoza, Spain},
        Ph.~Boucaud\address{Laboratoire de Physique Th\'eorique 
        et Hautes \'Energies, Univ. de Paris XI, 91405 Orsay, France},
        J.M.~Carmona\address{Dipartimento di Fisica, Universit\`a di
        Pisa, Via Buonarroti, 2, 56127 Pisa, Italy},
        J.L.~Cort\'es\num{a},
        J. Polonyi\address{Laboratoire de Physique Th\'eorique, 
        Univ. Louis Pasteur, 67084 Strasbourg Cedex, France, and \\
        Department of Atomic Physics, 
        L. E\"otv\"os University, Budapest, Hungary},   
        A.J.~van der Sijs\address{Swiss Center for Scientific Computing, 
        ETH Z\"urich, ETH-Zentrum, CH-8092 Z\"urich, Switzerland}}

\begin{document}
 
\begin{abstract}
We present a way of protecting a Dirac fermion interacting with a scalar
(Higgs) field from getting a mass from the vacuum. It is obtained
through an implementation of  
translational symmetry when the theory is formulated with a momentum cutoff,
which forbids the usual Yukawa term.
We consider that this mechanism can help to understand the
smallness of neutrino masses  without a tuning of the Yukawa coupling.
The prohibition of the Yukawa term for the neutrino 
forbids at the same time a gauge coupling  between the right-handed 
electron and neutrino.
We prove that this mechanism can be implemented on the lattice.
\end{abstract}

\maketitle

\section{HIGGS MECHANISM (SM)}

The Higgs mechanism is the mechanism to give mass to fermions and gauge
bosons in the Standard Model (SM). However, in the SM there are massless
fermions: the neutrinos.  In fact, a right-handed neutrino $\nu_R$ is not
introduced  so that it remains massless. However, most 
of the extensions of the SM imply the existence of a $\nu_R$. With the 
introduction of a $\nu_R$, the neutrino can be coupled to the Higgs field 
and get a Dirac mass $m_{\nu_e}(\bar \nu_{e_L} \nu_{e_R} + \mathrm{h.c.})$.
A fundamental problem is then 
to understand why $m_{\nu_e}/m_e$ is such a small number ($<10^{-5}$).
In the following we will give a possible answer to this problem by 
means of a mechanism
to protect a fermion coupled to the Higgs field from acquiring a mass from
the Higgs vacuum.

\section{MASS PROTECTION MECHANISM}

We present the following mechanism, based on two characteristics of the 
SM: first, the freedom in the choice of the representations of the symmetries
of the theory in which the elementary particles appear (we will consider 
specifically the translational symmetry) and second, that it is a low-energy
effective theory. This last fact implies the presence of a momentum cutoff
scale $\Lambda$. We will identify new representations of the translational
symmetry considering that the momentum cutoff $-\Lambda\leq p_\mu\leq\Lambda$
naturally reduces the Poincar\'e group of symmetry to a discrete subgroup:
in Euclidean space, that generated by rotations of $\pi/2$ in each plane, and
by translations of $\pi/\Lambda$ in each direction.

There are \textsl{only two} different representations of 
discretized translations ($r^\mu = n^\mu \pi/\Lambda$ ,
$n^\mu \in \Bbb{Z}$) which are \textsl{compatible with the usual 
representations of rotations:}
\begin{eqnarray}
\psi'(p) = e^{ir.p}\psi(p)&,&  \mathrm{and} \\
\psi'(p) = e^{ir.{\tilde p}}\psi(p)&,&  
{\tilde p}_\mu = p_\mu -\Lambda\, {\mathrm{sign}}(p_\mu).
\label{tilde}
\end{eqnarray}
In coordinate space, these representations are:
\begin{eqnarray}
\psi'(x)&=&\psi(x+r)\label{rep1}\\
\psi'(x)&=&e^{i\Lambda\sum_\mu r_\mu}\psi(x+r).\label{rep2}
\end{eqnarray}

To illustrate how the mass protection mechanism works, 
we will consider a chiral model
with a left and a right fermion coupled to a complex scalar field, and 
a different representation under translations for each chirality of the
fermion field. The physical interpretation of this will be that the two
chiralities are coupled differently to the physics beyond the cutoff.
Then, the usual Yukawa term in momentum space,
\begin{equation}
y \bar\psi_L([p+k])\phi(k)\psi_R(p),
\label{usualyukawa}
\end{equation}
is forbidden by translational invariance. In Eq.~(\ref{usualyukawa}),
$[p+k]$ is the momentum compatible with the cutoff obtained 
by adding or substracting if necessary $2\Lambda$ to the components 
of $p+k$. The interaction 
term compatible with the new implementation of translations is
\begin{equation} 
y \bar\psi_L(\widetilde{[p+k]})\phi(k)\psi_R(p),
\label{newyukawa}
\end{equation}
where the tilde symbol was already introduced in Eq.~(\ref{tilde}).
At leading order for the fermion propagator, one finds, in the 
case of the term~(\ref{usualyukawa}), a free fermion with mass 
$m=y\langle\phi\rangle$, and in the case of the term~(\ref{newyukawa}),
a massless fermion up to corrections proportional to inverse powers of the
cutoff $\Lambda$.

However, as the term~(\ref{newyukawa}) couples momentum modes that 
differ in $\Lambda$, a nonperturbative implementation of this 
mechanism could be problematic owing to the well-known fermion
doubling phenomenon.
Let us see that this is not the case.

\section{LATTICE IMPLEMENTATION}

On the lattice, we take for the  
representation of translations for the fermion field~: 
\begin{equation}
\psi'_{Lx}=e^{i\alpha_L}\psi_{L x+\hat\mu}\, ,\quad
\psi'_{Rx}=e^{i\alpha_R}\psi_{R x+\hat\mu},
\label{newreplat}
\end{equation}
under a translation of one lattice spacing 
in the $\hat\mu$ direction. As in the previous discussion, 
we will take $\alpha_L=0$ and $\alpha_R=\pi$ 
in order to have compatibility with the usual
representations of rotations.

The translational invariant lattice action is
\begin{equation}
S=S_{\mathrm{B}}+S_{\mathrm{F}}+S_{\mathrm{FB}},
\label{action}
\end{equation}
where
\begin{eqnarray}
S_{\mathrm{B}}&=&-\kappa\sum_{x,\mu}
(\phi^*_x\phi_{x+\hat\mu}+\phi^*_{x+\hat\mu}\phi_x) \nonumber \\
&&+\sum_x\{\phi^*_x\phi_x+
\lambda(\phi^*_x\phi_x-1)^2\} ,\\
S_{\mathrm{F}}&=&\sum_{x,\mu}\frac{1}{2}(\bar\psi_x\gamma_\mu\psi_{x+\hat\mu}-
\bar\psi_{x+\hat\mu}\gamma_\mu\psi_x), \label{kinetic} \\
S_{\mathrm{FB}}&=&y\sum_x (-1)^{\sum_\nu x_\nu}
(\bar\psi^{(1)}_{Lx}\phi_x\psi^{(1)}_{Rx} \nonumber \\
&&+\bar\psi^{(1)}_{Rx}\phi_x^*\psi^{(1)}_{Lx}), \label{newterm}
\end{eqnarray}
with
\begin{equation}
\psi^{(1)}(p)=F(p)\psi(p).
\end{equation}
$F(p)$ is a form factor required to be 1 for $p=0$ and
to vanish when $p$ equals any of the doubler momenta. 
With this method, we have a theory with 16 fermions, 15 of which do not
interact with physical particles and decouple from the real 
world~\cite{zara}. 

In order to do perturbation theory, let us set
\begin{equation}
\phi_x=\phi_{1x}+i\phi_{2x},
\end{equation}
and consider a scalar field $\phi$ with 
a VEV $\langle\phi_{1x}\rangle=v$, $\langle\phi_{2x}\rangle=0$.
Then we write
\begin{equation}
\phi_{1x}=v+\eta_{1x} \, , \quad \phi_{2x}=\eta_{2x},
\label{bosons}
\end{equation}
where $\eta_{1,2}$ represent the small perturbations.

Let us first note that the presence in the action of the 
term $S_{\mathrm{FB}}$ with such an unusual coupling does not modify the vacuum
$\langle\phi_{1x}\rangle=v$.  This is a consequence of both analytical and
numerical studies of the antiferromagnetic (AFM) phase of the chiral Yukawa
model~\cite{bock}. Under the change of variables 
$\phi'_x=\varepsilon_x\phi_x$, where $\varepsilon_x=(-1)^{\sum_\nu x_\nu}$, 
the action is invariant if the couplings are mapped according to
\begin{equation}
\left(\kappa,y\varepsilon_x\right)\longmapsto
\left(-\kappa,y\right).
\end{equation}
With these couplings, a stable AFM phase exists where the scalar gets a 
staggered mean value $\langle\phi'_{1x}\rangle=\varepsilon_x
v_{\mathrm{st}}$. We can then  conclude that the original vacuum
$\langle\phi_{1x}\rangle=v$ is  also a  stable  vacuum for the action
(\ref{action}).

In momentum space,
the inverse of the fermion propagator at tree-level order is
\begin{equation}
i\delta(p-p')\cruza s(p)+
 y v F(p)F_\pi(p)\delta(p-p'+\bbox{\pi}),
\end{equation}
where $\slash{s}(p)=\sum_\mu \gamma_\mu \sin p_\mu$, 
$F_\pi(p)\equiv F(p+\bbox{\pi})$, and $\bbox{\pi}\equiv(\pi,\pi,\pi,\pi)$. 
We have $F_\pi(0)=0.$
This matrix is
not diagonal in momentum space, as it connects $p$
with $p+\bbox{\pi}$ in a box of the form
\begin{equation}
G^{-1}(p)=\left(\begin{array}{cc}
i\cruza s(p) &  y v F(p)F_\pi(p) \\
 y v F_\pi(p)F(p) & i\cruza s(p+\bbox{\pi})
\end{array}\right)
\label{invprop}
\end{equation}
It can be diagonalized to give
\begin{equation}
G_{\mathrm{D}}=\frac{1}{i\cruza s(p)+im(p)}, \,\, m(p)= y v F(p)F_\pi(p).
\label{diagprop}
\end{equation}
This propagator has 16 poles at momenta $(0,0,0,0)$, $(\pi,\pi,\pi,\pi)$,
$(\pi,0,0,0)$, $(\pi,\pi,0,0)$, etc., which implies zero mass at tree level for
the physical fermion and all the doublers.
One can see through perturbative calculations and non-perturbative
arguments that this masslessness is maintained at every
loop order~\cite{hep-th}.

We are interested in the continuum limit of the theory because we want 
to apply it to energy scales $E\ll\Lambda$. In the limit $\Lambda\to\infty$,
the propagator~(\ref{diagprop}) becomes
$(-i)/\slash{p}$, that is, a massless fermion propagator. This limit
is well defined because it corresponds to the second order phase transition
of the AFM phase in the chiral Yukawa model, where we have restoration of
rotational invariance and renormalizability of the theory. In summary, we 
have obtained a massless fermion in the low-energy theory by using
transformation laws under the symmetries of the theory related to the 
presence of the scale $\Lambda$, that is, related to the properties of the
theory at the next level $E>\Lambda$.

\section{APPLICATION TO THE SM}

The present mechanism could be applied in the framework of the SM 
with $\nu_R$ to 
understand the absence of a neutrino Dirac mass, 
by simply choosing a different representation for the L and R
chiralities of this fermion under translational symmetry. As in the
SM $e_L$ and
$\nu_L$ are coupled by the gauge field, they should appear in the same
representation, together with $e_R$ (in order to have the usual Higgs
mechanism for the electron). Then the right-handed electron and the
right-handed neutrino are in different representations and they
cannot be in the same weak isospin multiplet. 
This situation is in fact assumed in the SM.

Recent oscillation results~\cite{results} suggest that the neutrino could
have a small mass. The usual explanation for this requires the introduction
of Majorana terms, which violate lepton number conservation. In the 
framework of the minimal SM as an effective theory, 
a Majorana mass can be generated
for the neutrino by the dimension five operator
\begin{equation}
\psi^{\mathrm{T} }_{L}(x) C \psi^{}_{L}(x)\phi(x)\phi(x).
\label{majorana}
\end{equation}
In the SM with $\nu_R$, the see-saw 
mechanism~\cite{yanagida} balances the Dirac term $\bar\psi_L(x)\psi_R(x)$ and
the Majorana term $\psi^{\mathrm{T} }_{R}(x) C \psi_{R}(x)$ in order to 
explain a small mass for the neutrino.

The mass protection mechanism proposed in this work allows a $\nu_R$ in
the theory \textsl{without the generation of a Dirac mass.} 
Also, a scenario of
almost-degenerate neutrinos (the relevant one in cosmology) could be explained
in an easier way after having eliminated the hierarchy of the Dirac mass
matrix. Besides that, if no Majorana terms are allowed in the model,
the neutrino oscillations could be due to effects of order
$v/\Lambda$, and compatible with lepton number conservation
in the framework of this mass protection mechanism.

\end{document}